\documentclass[aps,twocolumn,showpacs]{revtex4}
\usepackage{stmaryrd}
\usepackage{amsfonts}
\usepackage{mathrsfs}
\usepackage{amsmath}
\usepackage{amscd}
\usepackage{graphicx}
\usepackage{booktabs}

\begin{document}
\title{Experimental Realization of Nonadiabatic Holonomic Quantum Computation}
\author{ Guanru Feng$^{1,2}$, Guofu Xu$^{1,2}$ and Guilu Long$^{1,2}$}
\affiliation{$^1$ State Key Laboratory of Low-dimensional Quantum Physics and Department of Physics, Tsinghua University, Beijing 100084, China\\
         $^2$ Tsinghua National Laboratory of Information Science and Technology, Beijing 100084, China}
\begin{abstract}
Due to the geometric nature, holonomic quantum computation is
fault-tolerant against certain types of control errors. Although
proposed more than a decade ago, the experimental realization of
holonomic quantum computation is still an open challenge. In this
Letter,  we report the first experimental demonstration of
nonadiabatic holonomic quantum computation in a liquid NMR quantum
information processor. Two non-commuting one-qubit holonomic gates,
rotations about $x$- and $z$-axes, and the two-qubit holonomic CNOT
gate are realized by evolving the work qubits
and an ancillary qubit nonadiabatically. The successful realizations
of these universal elementary gates in nonadiabatic holonomic
quantum computation demonstrates the experimental feasibility of this quantum computing
paradigm.
\end{abstract}
\maketitle

{\noindent\it 1. Introduction.} Holonomic quantum computation (HQC)
was first proposed by Zanardi and Rasetti \cite{HQC}. In their
original work,  the twisting of eigenspaces of adiabatically varying
Hamiltonian  was used to manipulate quantum states in a robust
manner.  Due to the geometric nature,  HQC is robust against certain
types of control errors.  Since control errors are one main obstacle
to the realization of quantum computation, HQC has become one
promising quantum computation paradigm and attracted increasing
interests recently
\cite{Jones,Duan,Wang,Zhu,Wu,Cen,Zhang,Feng,Oreshkov,Golovach,Tong,Thomas,falci,leek,mottonen,pechal,Xu}.

Early HQC is based on adiabatic evolution, in which states are
encoded in degenerate eigenstates of a Hamiltonian, and gates are
accomplished by adiabatically varying the Hamiltonian along a loop
in the parameter space. Because of the adiabatic requirement, long
run-time is naturally required in the parametric control in
adiabatic HQC (AHQC). This not only limits the gate speed, but also
exposes the system to environment for a long time, and consequently
leads to decoherence and reduces the efficiency of AHQC. To
overcome these drawbacks in AHQC, nonadiabatic HQC (NHQC) has been
pursued, and several NHQC protocols have been proposed \cite{Tong,
Xu}. In NHQC, the long run-time requirement is avoided, while still
retaining all the robust advantages, making NHQC a very appealing
quantum computing paradigm.

In this Letter, we report the first experimental realization of NHQC
using a liquid NMR quantum information processor. The NHQC scheme we
realize is based on a variant of the recently proposed NHQC scheme
in Ref. \cite{Xu}.  In our modified NHQC scheme, decoherence-free subspace is not used and nonadiabatic
holonomic evolution is achieved by nonadiabatically evolving the
work qubits and an ancillary qubit. To experimentally realize universal quantum
computation, nonadiabatic one-qubit holonomic rotation gates about
$x$- and $z$-axes and the nonadiabatic two-qubit holonomic CNOT gate
are successfully implemented using a
three-qubit NMR quantum information processor. These
results demonstrate the experimental feasibility of NHQC.

{\noindent \it 2. Theoretical protocol.}  We first briefly review
the holonomic conditions. Consider an $N$-dimensional quantum system
with its Hamiltonian $H(t)$. Assume the state of the system is
initially in a $M$-dimensional subspace $\mathcal{S}(0)$ spanned by
a set of orthonormal basis vectors
$\{|\phi_{k}(0)\rangle\}_{k=1}^M$. It has been proved that
\cite{Tong,Xu} the evolution operator is a holonomic matrix acting
on $\mathcal{S}(0)$ if $|\phi_{k}(t)\rangle$ satisfy the following
conditions,
\begin{eqnarray}
& \textrm{(i)} & \ \ \sum_{k=1}^M |\phi_{k}(\tau)\rangle \langle
\phi_{k}(\tau)| =
\sum_{k=1}^M |\phi_{k}(0)\rangle \langle \phi_{k}(0)|, \label{conditions1}\\
& \textrm{(ii)} & \ \ \langle
\phi_{k}(t)|H(t)|\phi_{l}(t)\rangle=0,\ k,l=1,...,M,
\label{conditions}
\end{eqnarray}
where $\tau$ is the evolution period and
$|\phi_{k}(t)\rangle=\textbf{T}\exp(-i\int_{0}^{t}H(t_1)dt_1)|\phi_{k}(0)\rangle$,
$\textbf{T}$ being time ordering.

Now we construct the universal set of NHQC gates. For the
nonadiabatic one-qubit holonomic rotation gates, a two-qubit system
is used. We choose the logical qubit states as
$|0\rangle_L=|10\rangle$, $|1\rangle_L=|11\rangle$. By such a
design, all the information of the logical qubit is encoded in the work qubit (the second qubit), and the first qubit acts as an ancillary qubit. We design two types of Hamiltonians,
 $H_{1}(\phi_{1})$ and $H_{2}(\phi_{2})$, to respectively realize
 two non-commuting nonadiabatic one-qubit gates,
\begin{eqnarray}
H_{1}(\phi_{1})&= &\frac{1}{2} (a_1(X_{1}X_{2}+Y_{1}Y_{2})+b_1(X_{1}Y_{2}-Y_{1}X_{2}) \nonumber
\\ &&-a_1X_{1}(I_{2}-Z_{2})-b_1Y_{1}(I_{2}-Z_{2})),\label{ham1}\\
H_{2}(\phi_{2})&=&\frac{1}{2}
(a_2(Y_{1}X_{2}-X_{1}Y_{2})-b_2X_{1}(I_{2}-Z_{2})),
\end{eqnarray}
where $a_1=J_1\cos(\phi_{1}/2)$, $b_1=J_1\sin(\phi_{1}/2)$,
$a_2=J_2\sin(\phi_{2}/2)$,  $b_2=J_2\cos(\phi_{2}/2)$, $I$ is
one-qubit identity matrix, and $X$, $Y$, $Z$ are Pauli matrices. In
the basis $\{|00\rangle, |01\rangle, |10\rangle, |11\rangle\}$, the
evolution operators $U_{1}^{\phi_{1}}(\tau _{1})$ and
$U_{2}^{\phi_{2}}(\tau _{2})$ generated by $H_1(\phi_{1})$ and
$H_2(\phi_{2})$ respectively  read
\begin{align}
U_{1}^{\phi_{1}}(\tau _{1})=&\left(\begin{array}{cccc}
1 & 0 & 0 &0\\
0 & -1 & 0 &0\\
0 & 0 & 0 & e^{-i\phi_{1}}\\
0 & 0 & e^{i\phi_{1}} & 0\end{array}\right),
\end{align}
\begin{align}
U_{2}^{\phi_{2}}(\tau _{2})=&\left(\begin{array}{cccc}
1 & 0 & 0 &0\\
0 & -1 & 0 &0\\
0 & 0 & \cos \phi_{2} & i \sin \phi_{2}\\
0 & 0 & -i \sin \phi_{2} & -\cos
\phi_{2}\end{array}\right),\label{unitary2}
\end{align}
where $J_1\tau_{1}={\pi}/\sqrt{2}$ and $J_2\tau _{2}=\pi$. According
to Eqs. (\ref{ham1})-(\ref{unitary2}), it is readily to prove that
both conditions (i) and (ii) are satisfied if the state of the
two-qubit system is initially in the logical subspace
$S_1^{L}=\{|0\rangle_{L}, |1\rangle _{L}\}$. So
$U_{1}^{\phi_{1}}(\tau _{1})$ and $U_{2}^{\phi_{2}}(\tau _{2})$ are
holonomic matrices acting on $S_1^{L}$. In the basis
$\{|0\rangle_{L}, |1\rangle _{L}\}$, $U_{1}^{\phi_{1}}(\tau _{1})$
and $U_{2}^{\phi_{2}}(\tau _{2})$ are respectively equivalent to
\begin{eqnarray}
U_{xz}(\phi_{1})&=&\left(\begin{array}{cc}
0 & e^{-i\phi_{1}}\\
e^{i\phi_{1}} & 0\end{array}\right),\\
U_{zx}(\phi_{2})&=&\left(\begin{array}{cc}
\cos \phi_{2} & i \sin \phi_{2}\\
-i \sin \phi_{2} & -\cos \phi_{2}\end{array}\right).
\end{eqnarray}
Then the one-qubit holonomic rotation gates about $x$- and $z$-axes
acting on the space $S_1^{L}$ can be constructed by
 using  $U_{xz}$ and $U_{zx}$,
\begin{align}
&R_{z}^{L}(\theta)=U_{xz}(0)U_{xz}(-{\frac{\theta}{2}})\rightarrow U_{1}^{0}(\tau _{1})U_{1}^{-\frac{\theta}{2}}(\tau _{1}),\\
&R_{x}^{L}(\phi)=U_{zx}(0)U_{zx}(-{\frac{\phi}{2}})\rightarrow U_{2}^{0}(\tau _{2})U_{2}^{-\frac{\phi}{2}}(\tau _{2}).
\end{align}
From the above two gates, an arbitrary one-qubit NHQC operation can
be built.

The nontrivial two-qubit NHQC gate we realize is the nonadiabatic
holonomic CNOT gate. A three-qubit system is used to implement this
gate. $|100\rangle$, $|101\rangle$, $|110\rangle$ and $|111\rangle$
are encoded as $|00\rangle_{L}$, $|01\rangle_{L}$, $|10\rangle _{L}$
and $|11\rangle _{L}$. We see that all the information of the
logical two-qubit state is encoded in the two work qubits (the
second qubit and the third qubit), and the first qubit acts as an ancillary
qubit. The Hamiltonian $H_{3}$ for realizing the CNOT gate can be
expressed as
\begin{align}
H_{3}=&\frac{J_3}{4} (X_{1}(I_{2}-Z_{2})X_{3}+Y_{1}(I_{2}-Z_{2})Y_{3}\nonumber\\&-X_{1}(I_{2}-Z_{2})(I_{3}-Z_{3})).
\label{H3}
\end{align}
Letting the evolution time satisfy the condition $J_3\tau
_{3}={\pi/\sqrt{2}}$, the evolution operator in the basis $\{|000\rangle, |001\rangle, |010\rangle,
|011\rangle\,|100\rangle, |101\rangle, |110\rangle,
|111\rangle\}$ reads
\begin{align}
U_{3}(\tau _{3})={\rm Diag}[1,1,1,-1,1,1,{ X}].\label{unitary3}
\end{align}
According to Eqs. (\ref{H3}) and (\ref{unitary3}), we can prove that
both conditions (i) and (ii) are satisfied if the state of the
three-qubit system is initially in the logical subspace
$S_2^{L}=\{|00\rangle_{L}, |01\rangle_{L}, |10\rangle _{L},
|11\rangle _{L}\}$. So $U_{3}(\tau _{3})$ is a holonomic matrix
acting on $S_2^{L}$. In the basis $\{|00\rangle_{L}, |01\rangle_{L},
|10\rangle _{L}, |11\rangle _{L}\}$, $U_{3}(\tau _{3})$ is
equivalent to the nonadiabatic holonomic CNOT gate.

As the Hamiltonians $H_{1}(\phi_{1})$, $H_{2}(\phi_{2})$ and $H_{3}$ are time-independent, their holonomic
evolution operators can be respectively written as
\begin{align}
&U_{1}^{\phi_{1}}(\tau _{1})  = \Pi_{l=1}^{N_1} U_{1}^{\phi_{1}}(\Delta t_{1})
,\label{U1}\\
&U_{2}^{\phi_{2}}(\tau _{2})= \Pi_{l=1}^{N_{2}} U_{2}^{\phi_2}(\Delta t_{2})
,\label{U2}\\
&U_{3}(\tau _{3}) = \Pi_{l=1}^{N_{3}} U_{3}(\Delta t_{3}),\label{U3}
\end{align}
where $\Delta t_i$ ($i\in\{1,2,3\}$) is small time interval and its
value is ${\tau_{i}}/{N_{i}}$, $N_{i}$ being the number of the time
steps of the holonomic evolution. By using a modification of the
Trotter formula which is correct up to $(\Delta t)^2$
\cite{processtomography1}, the short time evolutions respectively
read
\begin{align}
U_{1}^{\phi_{1}}(\Delta t_{1})=&e^{-i\Delta t_1 H_{1}(\phi_{1})}\approx T_{1}^{\phi_1}(\Delta t_{1})\nonumber \\
=&e^{i\frac{\Delta t_1}{2}*\frac{b_1}{2}Y_{1}(I_{2}-Z_{2})}e^{i\frac{\Delta t_1}{2}*\frac{a_1}{2}X_{1}(I_{2}-Z_{2})}\nonumber\\&e^{-i\frac{\Delta t_1}{2}*\frac{b_1}{2}(X_{1}Y_{2}-Y_{1}X_{2})}e^{-i\Delta t_1*\frac{a_1}{2}(X_{1}X_{2}+Y_{1}Y_{2})}\nonumber\\&e^{-i\frac{\Delta t_1}{2}*\frac{b_1}{2}(X_{1}Y_{2}-Y_{1}X_{2})}e^{i\frac{\Delta t_1}{2}*\frac{a_1}{2}X_{1}(I_{2}-Z_{2})}\nonumber\\&e^{i\frac{\Delta t_1}{2}*\frac{b_1}{2}Y_{1}(I_{2}-Z_{2})}, \label{UU1}\\
U_{2}^{\phi_2}(\Delta t_{2})=&e^{-i\Delta t_2 H_{2}(\phi_{2})}\approx T_{2}^{\phi_2}(\Delta t_{2})\nonumber\\
 =&e^{i\frac{\Delta t_2}{2}*\frac{b_2}{2}X_{1}(I_{2}-Z_{2})}e^{-i\Delta t_2*\frac{a_2}{2}(Y_{1}X_{2}-X_{1}Y_{2})}\nonumber\\&e^{i\frac{\Delta t_2}{2}*\frac{b_2}{2}X_{1}(I_{2}-Z_{2})},\label{UU2}\\
U_{3}(\Delta t_{3})=&e^{-i\Delta t_3H_{3}}\approx T_{3}(\Delta t_{3}) \nonumber\\
= & e^{i\frac{\Delta t_3}{2}*\frac{J_{3}}{4}X_{1}(I_{2}-Z_{2})(I_{3}-Z_{3})}e^{-i\Delta t_3*\frac{J_{3}}{4}X_{1}(I_{2}-Z_{2})X_{3}}\nonumber\\&e^{-i\Delta t_3*\frac{J_{3}}{4}Y_{1}(I_{2}-Z_{2})Y_{3}}e^{i\frac{\Delta t_3}{2}*\frac{J_{3}}{4}X_{1}(I_{2}-Z_{2})(I_{3}-Z_{3})}.
\label{UU3}
\end{align}
Here $T_{1}^{\phi_1}(\Delta t_{1})$, $T_{2}^{\phi_2}(\Delta t_{2})$
and $T_{3}(\Delta t_{3})$ can be realized by a combination of
radio-frequency pulses and evolutions of the J-coupling constants
between the neighboring qubits in NMR technique
\cite{average1,average2,average3,PhysRevA.52.3457,apl76,3body,PhysRevA.78.012328}. According to
Eqs. (\ref{U1})-(\ref{UU3}), the nonadiabatic holonomic gates
$R_{z}^{L}(\theta)$, $R_{x}^{L}(\phi)$ and $U_{cnot}^L$ can be
realized  by
\begin{align}
R_{z}^{L}(\theta) \longrightarrow &\Pi_{l=1}^{N_{1}} T_{1}^{0}(\Delta t_{1})\Pi_{l=1}^{N_{1}} T_{1}^{-\frac{\theta}{2}}(\Delta t_{1}),\\
R_{x}^{L}(\phi) \longrightarrow &U_{2}^{0}(\tau _{2})\Pi_{l=1}^{N_{2}} T_{2}^{-\frac{\phi}{2}}(\Delta t_{2}),\label{rx}\\
U_{cnot}^L \longrightarrow&\Pi_{l=1}^{N_{3}} T_{3}(\Delta t_{3}).
\end{align}\\
Notably, in Eq. (\ref{rx}), $U_{2}^{0}(\tau _{2})$ can be implemented directly with no approximations, thus it reads $U_{2}^{0}(\tau _{2})=e^{i\frac{\tau_{2}J_{2}}{2}(X_{1}I_{2}-X_{1}Z_{2})}$.

\begin{figure}[!ht]
\centering
\includegraphics[width=3in,height=2.5in]{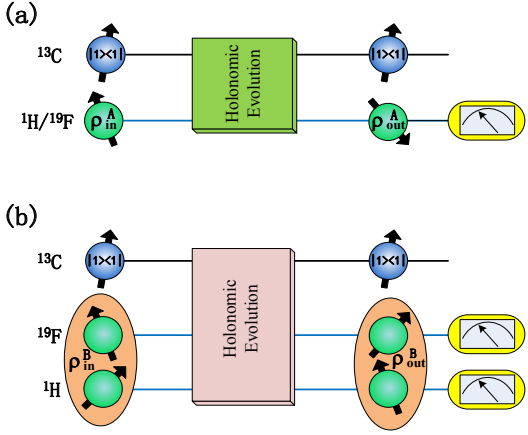}
\caption{(color online) Circuits for the NHQC gates. (a) The
one-qubit NHQC gates. (b) The two-qubit NHQC gate. Both in (a) and
(b), ${\rm ^{13}C}$ acts as an ancillary qubit and stays in state
$|1\rangle \langle 1|$ before and after the nonadiabatic holonomic
evolutions. ${\rm ^{19}F}$ and ${\rm ^{1}H}$ nuclear spins are the two work
qubits.} \label{curcirt}
\end{figure}

{\noindent \it 3. Experimental procedures and results.} Figures
\ref{curcirt} (a) and (b) respectively illustrate the
implementations of the nonadiabatic one-qubit and two-qubit
holonomic gates. The diethyl-fluoromalonate dissolved in $d6$
acetone is used as the NMR quantum processor. ${\rm ^{13}C}$, ${\rm ^{19}F}$ and
${\rm ^{1}H}$ nuclear spins respectively act as the ancillary qubit and
the two work qubits. Both in the one-qubit and two-qubit cases, the
ancillary qubit stays in state $|1\rangle \langle 1|$ before and
after the nonadiabatic holonomic evolutions. The input states
(output states) of the work qubits are denoted as $\rho_{in}^{A}$
($\rho_{out}^{A}$) and $\rho_{in}^{B}$ ($\rho_{out}^{B}$) for the
one-qubit and two-qubit cases respectively.

\begin{figure}[!ht]
\centering
\includegraphics[width=3.3in,height=2in]{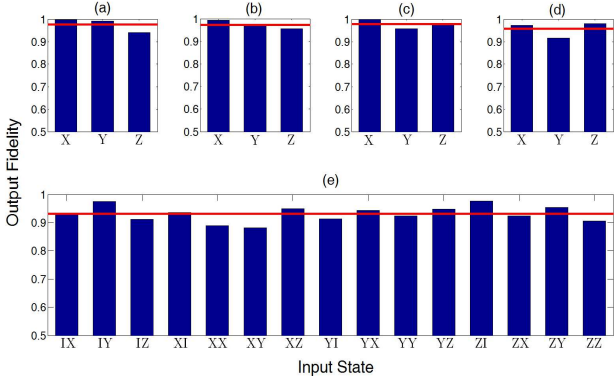}
\caption{(color online) The experimental unattenuated output state fidelities for
the NHQC gates. In (a), (b), (c) and (d) are the fidelities of
$\rho_{out}^{A}$ for $R_{z}^{L}({\pi}/{2})$, $R_{z}^{L}(\pi)$,
$R_{x}^{L}({\pi}/{2})$ and $R_{x}^{L}(\pi)$ respectively, applied to
input states $X$, $Y$ and $Z$. $R_{z}^{L}({\pi}/{2})$ and
$R_{x}^{L}({\pi}/{2})$ are implemented on ${\rm ^{19}F}$; $R_{z}^{L}(\pi)$
and $R_{x}^{L}(\pi)$ are implemented on ${\rm ^{1}H}$. In (e) are the
fidelities of $\rho_{out}^{B}$ for $U_{cnot}^L$, applied to 15
different input states listed in Eq. (\ref{input}).  The average
fidelities (the red solid horizontal lines) are 97.6\%, 97.3\%, 97.9\%, 95.7\% and
93.12\% in (a)-(e), respectively. } \label{fidelity}
\end{figure}

\begin{figure}[!ht]
\includegraphics[width=3.5in,height=3in]{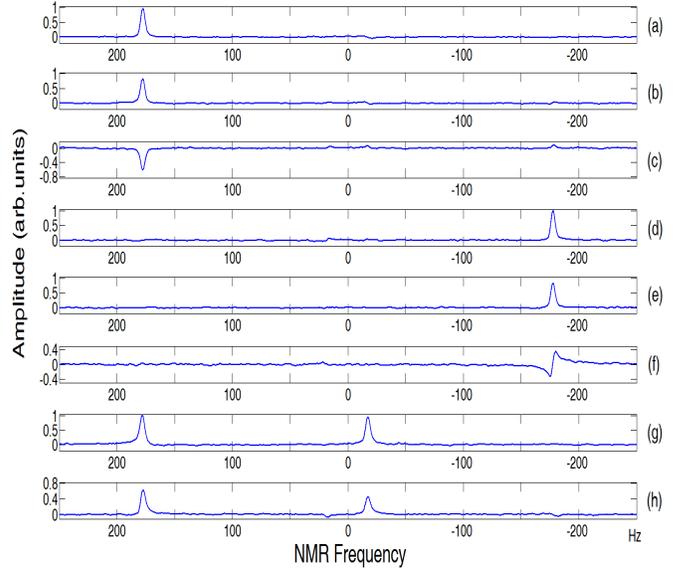}
\centering
\caption{(color online) Experimental spectra of ${\rm ^{13}C}$. (a) and
(d) are respectively the spectra obtained by observing the states
with ${\rm ^{1}H}$ and ${\rm ^{19}F}$ in $\rho_{in}^{A}=X$, with no holonomic
operations. (b), (c), (e) and (f) are the spectra of
$\rho_{out}^{A}$, starting with the initial states
$\rho_{in}^{A}$=$X$, and applying the holonomic operations
$R_{x}^{L}(\pi)$, $R_{z}^{L}(\pi)$, $R_{x}^{L}({\pi}/{2})$ and
$R_{z}^{L}({\pi}/{2})$, respectively. $R_{x}^{L}(\pi)$ and
$R_{z}^{L}(\pi)$ are implemented on ${\rm ^{1}H}$; $R_{x}^{L}({\pi}/{2})$
and $R_{z}^{L}({\pi}/{2})$ are implemented on ${\rm ^{19}F}$. (g) is the
spectrum of the initial state $\rho_{in}^{B}=IX$. (h) is the spectrum of  $\rho_{out}^{B}$ after the holonomic operation $U_{cnot}^L$ with the
initial state $\rho_{in}^{B}=IX$. (a), (d) and (g) are used as
reference spectra, to which (b), (c), (e), (f) and (h) are
normalized. All the observation is realized by transfering the
states of the work qubits to ${\rm ^{13}C}$ and then observing ${\rm ^{13}C}$.}
\label{spectra}
\end{figure}

We here realize the following four one-qubit NHQC gates,
$R_{z}^{L}({\pi}/{2})$, $R_{z}^{L}(\pi)$, $R_{x}^{L}({\pi}/{2})$,
$R_{x}^{L}(\pi)$, and the NHQC CNOT gate $U_{cnot}^L$. In order to
demonstrate we can implement one-qubit NHQC gates on both ${\rm ^{19}F}$
and ${\rm ^{1}H}$, $R_{z}^{L}({\pi}/{2})$ and $R_{x}^{L}({\pi}/{2})$ are
implemented on ${\rm ^{19}F}$ and $R_{z}^{L}(\pi)$ and $R_{x}^{L}(\pi)$
are implemented on ${\rm ^{1}H}$. In our experiments, the number of
iterations are chosen to be $N_{1}=3$, $N_{2}=2$ and $N_{3}=2$. We
prepare the initial states using the cat-state method
\cite{pps,natcomm,natcomm1}. For the one-qubit gates, we prepare the
work qubit in $\rho_{in}^A$ and the ancillary qubit in $|1\rangle
\langle 1|$. Without loss of generality, the spectator work qubit is
prepared in $|0\rangle \langle 0|$. Specifically, the NMR processor
is initialized in the pseudopure states $|1\rangle \langle 1|
\otimes \rho_{in}^{A} \otimes |0\rangle \langle 0| $ (for
$R_{z}^{L}({\pi}/{2})$ and $R_{x}^{L}({\pi}/{2})$) or $|1\rangle
\langle 1| \otimes |0\rangle \langle 0| \otimes \rho_{in}^{A} $ (for
$R_{z}^{L}(\pi)$ and $R_{x}^{L}(\pi)$). For the CNOT gate, the
ancillary qubit is also prepared in $|1\rangle \langle 1|$ and the
whole state of the NMR processor is $|1\rangle \langle 1| \otimes
\rho_{in}^{B}$. In terms of the deviation matrices \cite{Chuang},
the input states $\rho_{in}^{A}$ and $\rho_{in}^{B}$ are prepared in
each of the following sets
\begin{align}
\rho_{in}^{A}\in \{X,Y,Z\}, \label{input1}
\end{align}
\begin{align}
\rho_{in}^{B}\in
&\{IX,IY,IZ,XI,XX,XY,XZ,YI,\nonumber\\&YX,YY,YZ,ZI,ZX,ZY,ZZ\}.
\label{input}
\end{align}
The output states $\rho_{out}^{A}$ and $\rho_{out}^{B}$ are
determined by quantum state tomography (QST) \cite{tomography1}.
To measure the sameness of the theoretical output state $\rho_{th}$ and the
experimental output state $\rho_{out}$, the attenuated and unattenuated state fidelities \cite{Fortunato,Weinstein}, which are respectively defined as ${\rm Tr}(\rho_{out}\rho_{th})/\sqrt{{\rm Tr} (\rho_{th}\rho_{th}){\rm Tr}(\rho_{in}\rho_{in})}$ and ${\rm Tr}(\rho_{out}\rho_{th})/\sqrt{{\rm Tr}(\rho_{out}\rho_{out}){\rm Tr}(\rho_{th}\rho_{th})}$, are used. The attenuated fidelity takes into account the signal loss, while the unattenuated fidelity ignores certain errors due to the signal loss and quantifies how similar in direction $\rho_{out}$ and $\rho_{th}$ are \cite{Fortunato,Weinstein}. The average experimental attenuated fidelities are
60.7\%, 61.8\%, 86.1\%, 77.7\%, 47.9\%  for the output states of  $R_z^L(\pi/2)$, $R_z^L(\pi)$, $R_x^L(\pi/2)$, $R_x^L(\pi)$ and $U_{cnot}^L$ respectively, while their average experimental unattenuated fidelities are 97.6\%, 97.3\%, 97.9\%, 95.7\%
and 93.12\% respectively. These numbers are in-line with the results of other experiments done using diethyl-fluoromalonate \cite{xinhuapeng1n}. The differences between the attenuated and unattenuated fidelities are consistent with the signal loss rates measured in our experiments (see Supporting Material). 
Figure \ref{fidelity} shows the unattenuated output state fidelities in our expriments. Figure \ref{spectra} shows example NMR
experimental spectra. Figure \ref{spectra} (a) (Fig. \ref{spectra}
(d)) is the ${\rm ^{13}C}$ spectrum of the input state $\rho_{in}^{A}=X$
for ${\rm ^{1}H}$ (${\rm ^{19}F}$), with ${\rm ^{19}F}$ (${\rm ^{1}H}$) in state
$|0\rangle\langle 0|$. Figures \ref{spectra} (b), (c), (e), (f) show
the ${\rm ^{13}C}$ spectra of the output states $\rho_{out}^{A}$, after
implementing $R_{x}^{L}(\pi)$, $R_{z}^{L}(\pi)$,
$R_{x}^{L}({\pi}/{2})$ and $R_{z}^{L}({\pi}/{2})$ to the input
states $\rho_{in}^{A}=X$, respectively. Figures \ref{spectra} (g),
(h) show the spectra of $\rho_{in}^{B}=IX$ and $\rho_{out}^{B}$
after applying $U_{cnot}^L$.

\begin{figure}[!ht]
\centering
\includegraphics[width=3.5in,height=2.8in]{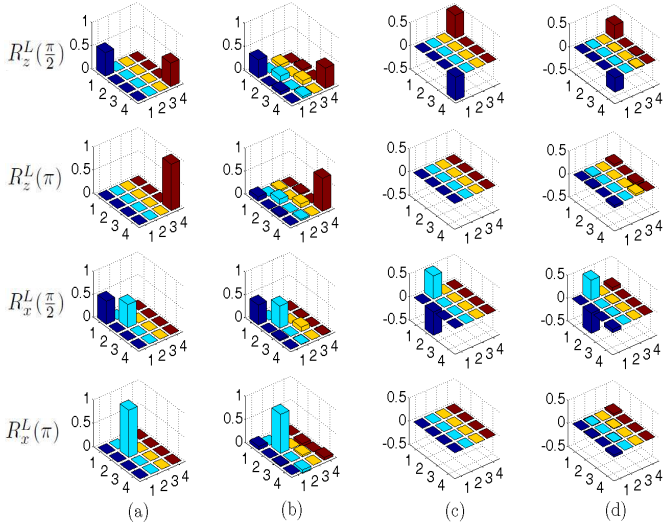}
\caption{(color online) The QPT $\chi$ matrices of one-qubit holonomic gates $R_{z}^{L}(\frac{\pi}{2})$, $R_{z}^{L}(\pi)$, $R_{x}^{L}(\frac{\pi}{2})$ and $R_{x}^{L}(\pi)$. The (a) and (c) columns are the real parts and imaginary parts of the theoretical $\chi$ matrices, respectively. The (b) and (d) columns are the real parts and  imaginary parts of the experimental $\chi$ matrices, respectively. The numbers in the $x$- and $y$- axes refer to the operators  in the operator basis set $\{I,X,-iY,Z\}$.}
\label{single}
\end{figure}

\begin{figure}[!ht]
\centering
\includegraphics[width=3.5in,height=2.5in]{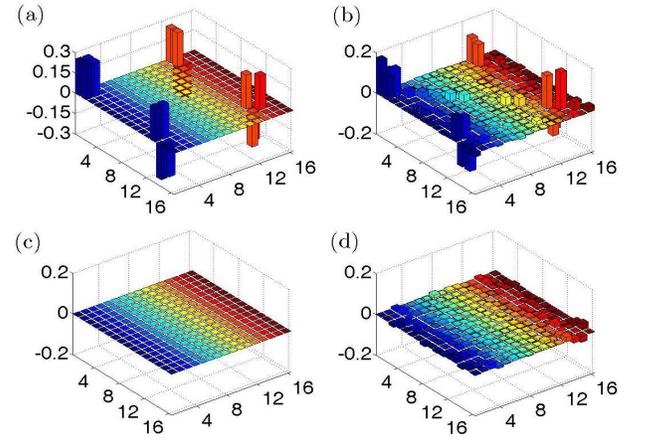}
\caption{(color online) The QPT $\chi$ matrices of $U_{cnot}^L$. (a) and (c) are the
real part and imaginary part of the theoretical $\chi$ matrix,
respectively. (b) and (d) are the real part and imaginary part of
the experimental $\chi$ matrix, respectively. The numbers 1 to 16 in the $x$- and $y$- axes
refer to the operators in the operator basis set  $\{II,IX,-iIY,IZ,XI,XX,-iXY,XZ,-iYI,-iYX,$
$-YY,-iYZ,ZI,ZX,-iZY,ZZ\}$.}
\label{double}
\end{figure}

Quantum process tomography (QPT) \cite{processtomography} is used to
quantitatively describe the implementation of the NHQC gates. According to QPT, each quantum process is characterized by a $\chi$ matrix. For a given
input state $\rho_{in}$, the output state is expressed as
$\rho_{out}=\Sigma_{k,l}\chi _{kl}e_{k}\rho_{in}e_{l}^{\dagger}$,
where $e_{k}$ belongs to an operation basis set. The elements of the operator basis set for the one-qubit
and two-qubit cases can be respectively chosen as
\begin{align}
e_{k}\in \{I,X,-iY,Z\},k=1,...,4, \label{singlebasis}
\end{align}
\begin{align}
e_{k}\in
\{&II,IX,-iIY,IZ,XI,XX,-iXY,XZ,-iYI,\nonumber\\&-iYX,-YY,-iYZ,ZI,ZX,-iZY,ZZ\},\nonumber\\&k=1,...,16.
\label{doublebasis}
\end{align}
The QPT $\chi$ matrix is calculated using the output states via the
technique described in Ref. \cite{processtomography1}. The
experimental $\chi$ for one-qubit and two-qubit gates are shown in
Figs. \ref{single} and \ref{double} respectively. We use $\chi$ fidelities to evaluate the performance of NHQC gates. The attenuated $\chi$ fidelities 
$|{\rm Tr}(\chi_{exp}\chi_{th}^\dagger)|$ \cite{Weinstein}, which take into account the signal loss,  are
 70.5\%, 71.3\%, 89.5\%, 83.3\% and 51.2\% for the $R_{z}^{L}({\pi}/{2})$,
$R_{z}^{L}(\pi)$, $R_{x}^{L}({\pi}/{2})$, $R_{x}^{L}(\pi)$ and
$U_{cnot}^L$ gates respectively. The deviations between $\chi_{th}$ and $\chi_{exp}$ are mainly caused by overall loss of signal. To see the
sameness of theoretical and experimental quantum processes when ignoring certain errors due to signal loss, we use the unattenuated $\chi$ fidelity  defined as $|{\rm Tr}(\chi_{exp}\chi_{th}^\dagger)|/\sqrt{{\rm Tr}(\chi_{exp}\chi_{exp}^{\dagger}){\rm Tr}(\chi_{th}\chi_{th}^\dagger)}$ \cite{Weinstein,fidelity,fidelity2}. 
The unattenuated experimental $\chi$ fidelities of the gates $R_{z}^{L}({\pi}/{2})$,
$R_{z}^{L}(\pi)$, $R_{x}^{L}({\pi}/{2})$, $R_{x}^{L}(\pi)$ and
$U_{cnot}^L$ are $95.9\%$, $95.9\%$, $98.1\%$, $96.3\%$ and
$91.43\%$, respectively. It is interesting to note that the Trotter
approximations in Eqs. (\ref{UU1}), (\ref{UU2}) and (\ref{UU3}) give
very good approximations to the exact evolution and the theoretical $\chi$ fidelities are $99.2\%$, $98.6\%$, $99.2\%$, $97.4\%$ and $98.7\%$ for
$R_{z}^{L}({\pi}/{2})$, $R_{z}^{L}(\pi)$, $R_{x}^{L}({\pi}/{2})$,
$R_{x}^{L}(\pi)$ and $U_{cnot}^L$ respectively.

{\noindent \it 4.  Summary.} As a proof of principle, we
experimentally implemented NHQC via a NMR quantum information
processor using a variant version of the scheme proposed in Ref.
\cite{Xu}. In our experiments, one-qubit nonadiabatic holonomic
gates and two-qubit holonomic CNOT gate, which compose a universal
set of NHQC gates, are implemented by using an
ancillary qubit which provides the additional dimension needed in
the holonomic evolution. This is the first experimental
demonstration of NHQC, which is a step towards
fault-tolerant quantum computing. The successful realizations of
these universal elementary gates in NHQC demonstrate the feasibility of implementing NHQC using present experimental techniques.

{\bf Acknowledgments} This work is supported by the National Natural
Science Foundation of China(Grant Nos. 11175094, 91221205), and the
National Basic Research Program of China (2009CB929402,
2011CB9216002). Thanks to IQC, University
of Waterloo, for providing the NMR software compiler.\\

\space

\end{document}